\documentclass[12pt]{article}

\usepackage{theorem,amssymb,amsbsy,latexsym}

\theoremstyle{plain}

{\newtheorem{Lem}{Lemma}}
{\newtheorem{Prop}{Proposition}}

{\theorembodyfont{\upshape}\newtheorem{Rem}{Remark}}

\newcommand{\mbf}[1]{{\boldsymbol {#1} }}

\newcommand{\ga}{\gamma} 
\newcommand{\ee}[1]{{\rm e}^{#1}}
\newcommand{\ii}{{\rm i}}
\newcommand{\dd}{{\rm d}}
\newcommand{\f}{\frac}

\newcommand{\vx}{{\bf x}}
\newcommand{\vy}{{\bf y}}
\newcommand{\vz}{{\bf z}}
\newcommand{\vP}{{\bf P}}
\newcommand{\vn}{{\bf n}}
\newcommand{\vm}{{\bf m}}

\newcommand{\vE}{{\bf E}}
\newcommand{\vmu}{\underline{{\mbf \mu}}}
\newcommand{\vzero}{\underline{{\mbf 0}}}

\newcommand{\Ref}[1]{(\ref{#1})}
\newcommand{\binom}[2]{\left(\!\!\bma{c} {#1}\\ {#2}\ema\!\!\right)}

\newcommand{\eps}{\varepsilon}

\newcommand{\half}{\mbox{$\frac{1}{2}$}}

\newcommand{\R}{{\mathbb R}}
\newcommand{\C}{{\mathbb C}}
\newcommand{\Z}{{\mathbb Z}}

\newcommand{\N}{{\mathbb N}}

\newcommand{\cC}{{\cal C}}

\newcommand{\cP}{{\cal P}}

\newcommand{\QED}{\hfill$\square$}

\newcommand{\eq}{\begin{equation}}
\newcommand{\eqend}{\end{equation}}
\newcommand{\eqa}{\begin{eqnarray}}
\newcommand{\nonueqa}{\begin{eqnarray*}}
\newcommand{\eqaend}{\end{eqnarray}}
\newcommand{\nonueqaend}{\end{eqnarray*}}
\newcommand{\nonu}{\nonumber \\ \nopagebreak}
\newcommand{\bma}[1]{\begin{array}{#1}}
\newcommand{\ema}{\end{array}}
\newcommand{\bc}{\begin{center}}
\newcommand{\ec}{\end{center}}

\newcounter{saveeqn}
\newcounter{App} 
\newcommand{\app}{%
\stepcounter{App}%
\setcounter{saveeqn}{\value{equation}}%
\setcounter{equation}{0}%
\renewcommand{\theequation}{\Alph{App}\arabic{equation}} }

\begin{document}
\begin{flushright}
\today
\end{flushright}
\vspace{.4cm}

\begin{center}

{\Large \bf Algorithms to solve the Sutherland model}
\vspace{1 cm}

{\large Edwin Langmann}\\
\vspace{0.3 cm} {\em Theoretical Physics, Royal Institute of
Technology, SE-10044, Stockholm, Sweden}\\

\end{center}

\begin{abstract}
We give a self-contained presentation and comparison of two different
algorithms to explicitly solve quantum many body models of
indistinguishable particles moving on a circle and interacting with
two-body potentials of $1/\sin^2$-type. The first algorithm is due to
Sutherland and well-known; the second one is a limiting case of a
novel algorithm to solve the elliptic generalization of the Sutherland
model. These two algorithms are different in several details. We show
that they are equivalent, i.e., they yield the same solution and are
equally simple.
\smallskip 

\noindent {PACS:  02.30.Ik, 03.65.-w, 05.30.Pr\\
MSC-class: 81Q05, 35Q58}
\end{abstract}

\section{Introduction}

We recently presented a novel algorithm to solve the quantum version
of the elliptic Calogero-Moser-Sutherland system \cite{EL0,EL1}. In
the trigonometric limit, such an algorithm was discovered already
about thirty years ago by Sutherland \cite{Su1,Su2}. Somewhat
surprisingly, the former algorithm in that limit reduces to one which
is different from Sutherland's, even though it yields the same
solution and is equally simple. The purpose of this paper is to give a
detailed and self-contained comparison of these two algorithms,
including a proof of their equivalence.

The Sutherland model is defined by the differential operator 
\eq
\label{CS}
H = - \sum_{j=1}^N\frac{\partial^2}{\partial x_j^2} \; + \; 2 \lambda
(\lambda-1)\!\! \sum_{1\leq j<k\leq N} V(x_j-x_k) 
\eqend
with $-\pi \leq x_j \leq \pi$, $N=2,3,\ldots$, $\lambda>0$, and 
\eq 
\label{V}
V(r) = \f{1}{ 4 \sin^2(r/2) } . 
\eqend 
This differential operator defines a self-adjoint operator on the
Hilbert space of square integrable functions on $[-\pi,\pi]^N$,
providing a quantum mechanical model for $N$ indistinguishable
particles moving on a circle of length\footnote{To ease notation, we
set the length of space to $2\pi$ from the start. Of course, an
arbitrary length $L>0$ can be easily introduced by rescaling $x_j\to
(2\pi/L)x_j$, $H\to (2\pi/L)^2 H$, etc.} $2\pi$ and interacting with a
two body potential proportional to $V(r)$ where $\lambda$ determines
the coupling strength. (To be precise: This model corresponds to a
particularly nice self-adjoint extension of this differential operator
which, for $\lambda>1$, corresponds to the Friedrich's extension
\cite{RS2}.) To solve this model amounts to constructing a complete
set of eigenfunctions and corresponding eigenvalues of this
Hamiltonian.

The starting point for Sutherland's algorithm is the following 

\smallskip 

\noindent {\bf Fact 1 \cite{Su1}:} {\it The wave function\footnote{To
fix the phase of $\Psi_0$ unambiguously one can interpret
$\sin(r/2)^\lambda$ as $\lim_{\eps\downarrow 0} \sin(r/2+\ii
\eps)^\lambda$, for example.  Anyway, the phase ambiguities associated
with the exponentiated sines are irrelevant here. In Appendix~B.1 we
will have to be more careful about similar phase ambiguities in the
functions $F(\vx;\vy)$ defined below.}
\eq
\label{Psi}
\Psi_0(\vx) = \prod_{1\leq j<k\leq N}\psi(x_k-x_j)^\lambda 
\eqend
with
\eq
\label{psi}
\psi(r) = \sin(r/2) 
\eqend
is the ground state of the Sutherland Hamiltonian, $H\Psi_0=E_0\Psi_0$.}

Exploiting this fact, Sutherland constructed all other eigenfunctions
$f$ using the following ansatz,
\eq f(\vx) = \Psi_0(\vx) \Phi(\vx) \eqend 
where $\Phi$ are symmetric polynomials (i.e.\ non-negative powers) in
the variables $z_j=\exp( \ii x_j)$ \cite{Su2}.  The symmetric
polynomials thus obtained are the so-called Jack polynomials which
have been studied extensively in the mathematics literature, see e.g.\
\cite{McD,St}.

Our algorithm is based on the following

\smallskip 

\noindent {\bf Fact 2 \cite{EL0}:} {\it The function
\eq
\label{F}
 F(\vx;\vy) = \f{ \prod_{1\leq j<k\leq N} \psi(x_{k}-x_j)^{\lambda} 
\prod_{1\leq j<k\leq N} 
\psi(y_{j}-y_{k})^{\lambda}}{\prod_{j,k=1}^N
\psi(x_j-y_k)^{\lambda}} ,   
\eqend
$\psi(r)$ as in Eq.\ \Ref{psi}, obeys the following identity, 
\eq
\label{rem}
\sum_{j=1}^N\biggl(\f{\partial^2}{\partial x_j^2} -  
\f{\partial^2}{\partial y_j^2}  \biggr)F(\vx;\vy) 
=  2\lambda(\lambda-1)\sum_{1\leq j<k\leq N}\biggl( 
V(x_k-x_j) - V(y_j-y_k) \biggr)F(\vx;\vy)  
\eqend
with $V(r)$ as in Eq.\ \Ref{V}. 
}

\smallskip

Note that we can write this latter identity as 
\eq
\label{rem1}
H(\vx)F(\vx;\vy) = H(\vy)F(\vx;\vy) 
\eqend 
where $H$ is the differential operator in Eq.\ \Ref{CS} but acting on
different arguments $\vx$ and $\vy$, as indicated. The idea of our
algorithm is to take the Fourier transform of Eq.\ \Ref{rem1} with
respect to the variables $\vy$, and this yields an identity allowing
to construct eigenfunctions and the corresponding eigenvalues
(Proposition \ref{P1}).

It is interesting to note that Fact~2 holds true in the elliptic case
as well (in this case, $\psi(r)$ is a Jacobi Theta function
$\vartheta_1(r/2)$ with {\it nome} $q=\exp(-\beta/2)$ and $V(r)$ is
Weierstrass' elliptic function $\wp(r)$ with periods $2\pi$ and
$\ii\beta$) \cite{EL0}, in contrast to Fact~1 \cite{Su3}.  For the
convenience of the reader, an elementary proof of Fact~2 (in the
trigonometric case) is given in Appendix~A. (This proof uses Fact~1; a
self-contained proof valid also in the elliptic case will be given in
\cite{EL1}.)

The plan of the rest of this paper is as follows. In Section~2 we
review the Sutherland algorithm \cite{Su2}, mainly to introduce our
notation. Section~3 contains a detailed description of our
algorithm. In the final Section~4 we give the arguments which prove
that both algorithms are equivalent, despite of various differences.
Lengthy proofs are deferred to two Appendices.

\section{Sutherland's algorithm}
We use $H\Psi_0=E_0\Psi_0$ with \cite{Su1}
\eq
\label{E0}
E_0 = \f{1}{12}\lambda^2 N(N^2-1) 
\eqend
and make the ansatz $f=\Phi\Psi_0$. With that the eigenvalue equation 
$Hf=Ef$ becomes $H'\Phi=E'\Phi$ with $E'=E-E_0$ and \cite{Su2}
\eq
H' =  - \sum_{j=1}^N\frac{\partial^2}{\partial x_j^2} 
\; - \;   
\ii \lambda \sum_{1\leq j<k\leq N} \biggl( 
\f{ \ee{\ii x_j}+ \ee{\ii x_k}  }{\ee{\ii x_j} - \ee{\ii x_k}}
\biggr)
\biggl( \frac{\partial}{\partial x_j} - 
\frac{\partial}{\partial x_k} 
\biggr) \: . 
\eqend 
One now determines the action of $H'$ on symmetric polynomials
\eq
\label{Sn}
S_{\vn}(\vx) = \sum_P \prod_{j=1}^N \ee{\ii n_j x_{Pj} }
\eqend
where\footnote{Note that $n_j$ in Ref.\ \cite{Su2} corresponds to
$n_{N+1-j}$ here.} 
\eq
\label{n1n2}
n_1\geq n_2\geq \ldots \geq n_N \geq 0 
\eqend
and the sum is over all permutations $P$ of $\{1,2,\ldots,N\}$. 
Using the identity \cite{Su2}
$$
(\ee{\ii x} + \ee{\ii y}) 
\biggl( \f{ \ee{\ii  k x} - \ee{\ii k y}  }{\ee{\ii x} - \ee{\ii y}}\biggr)
= \ee{\ii kx} + \ee{\ii ky} + 2\sum_{\nu=1}^{k-1} 
\ee{\ii [(k-\nu)x +\nu y]}
$$
for $k>0$, one obtains
\eqa
H' S_{\vn} = E_{\vn}'S_{\vn} +  \lambda \sum_{1\leq j<k\leq N}
(n_j-n_k) \sum_{\nu=1}^{n_j-n_k-1}S_{\vn - \nu \vE_{jk}} 
\eqaend
where we introduced the notation
\eq
\label{vE}
(\vE_{jk})_\ell = \delta_{j\ell} -\delta_{k\ell} , 
\quad j,k,\ell = 1,2,\ldots, N 
\eqend
and defined
\eq
\label{Enp}
E_{\vn}' = \sum_{j=1}^N n_j^2 +  \lambda \sum_{1\leq j<k\leq N}(n_j-n_k)
= \sum_{j=1}^N \Bigl( n_j^2 +  \lambda [N+1-2j] n_j \Bigr)  \: . 
\eqend
We now introduce the notation 
\eq
\label{vmu}
\vmu = \sum_{1\leq j<k\leq N} \mu_{jk}\vE_{jk}
\eqend
for non-negative integers $\vmu_{jk}$, 
and observe that there is a natural order
on the set of all these $\vmu$ (which we can identify with 
$\N_0^{N(N-1)/2}$), 
\eq
\label{order}
\vmu<\vmu' \; \mbox{  if  }
\mu_{jk}<\mu_{jk}'\;   \mbox{ for all $j<k$.}  
\eqend
It is obvious that $H'S_{\vn}$ is a
finite linear combination of symmetrized plane waves $S_{\vn-\vmu}$
with $\vmu\geq \vzero$. We thus can make the following ansatz for the
eigenfunctions of $H'$,
\eq
\label{Phin}
\Phi_{\vn} = \sum_{\vmu\geq \vzero} c_{\vmu} 
S_{\vn-\vmu} 
\eqend
with 
\eq
c_{\vmu}=0  \quad \mbox{if $(\vn -\vmu)_j < (\vn - \vmu)_k$ 
for at least one $j<k$}.
\eqend
The latter condition shows that there are only a finite number of
non-zero $c_{\vmu}$, i.e., the $\Phi_{\vn}$ are polynomials. 
Then $H'\Phi_{\vn}=E'\Phi_{\vn}$ implies $E'=E_{\vn}'$ and
the following recursion relations for the coefficients $c_{\vmu}$, 
\eq
\label{bn}
\bigl[E_{\vn}'-E_{\vn - \vmu}'\bigr] c_{\vmu} =
\lambda \sum_{1\leq j<k\leq N}\sum_{\nu=1}^{n_j-n_k}
\bigl[(\vn - \vmu)_j - 
(\vn - \vmu )_k + 2\nu \bigr] 
c_{\vmu - \nu \vE_{jk} } 
\eqend
(we used the fact that the functions $S_{\vn}$ are linearly independent). 
We can set $c_{\vzero}=1$ (this fixes the normalization of the
eigenfunctions) and then determine the other $c_{\vmu}$ recursively,
which is possible provided that there is no resonance, i.e., if
$E_{\vn}'-E_{\vn+\vmu}'$ is non-zero. This is the case: we
shall prove at the end of this Section that 
\eq
\label{nores}
E_{\vn}'-E_{\vn-\vmu}'  
= \sum_{j<k}\mu_{jk} [(\vn-\vmu)_j - (\vn -\vmu)_k  + (n_j-n_k) +
2\lambda(k-j)] 
\eqend 
which is manifestly positive and shows that resonances indeed do not
occur. This completes the construction of eigenfunctions and
eigenvalues of the Sutherland model: Note that the symmetrized plane
waves $S_{\vn}$ provide a complete orthonormal basis of the
corresponding non-interacting Hamiltonian (obtained by setting
$\lambda=0$), and we have constructed eigenfunctions $f_{\vn}
=\Phi_{\vn}\Psi_0$ and corresponding eigenvalues $E_{\vn} =
E_{\vn}'+E_0$ with $E_0$ in Eq.\ \Ref{E0}, which are one-to-one to
this free solution which is known to provide a complete basis.

\begin{Rem} 
We can write (cf.\ Eq.\ \Ref{Enp}) $E_{\vn}' =
\sum_j [n_j+ \half \lambda(N+1-2 j )]^2 -E_0'$ with 
\eq E_0' =
\sum_{j=1}^N \f{1}{4}\lambda^2 (N+1-2j)^2 .  
\eqend 
It is easy to show
that $E_0'=E_0$ (cf.\ Eq.\ \Ref{E0}), which is somewhat remarkable
and implies the following simple form of the eigenvalues,
\eq 
\label{En}
E_{\vn} = \sum_{j=1}^N \Bigl( n_j + \lambda[\half(N+1)-j] \Bigr)^2 .
\eqend 
The novel algorithm in the next Section will yield this simple
form of the eigenvalues directly. 
\end{Rem}

For the convenience of the reader, we conclude this Section with a

\smallskip 

\noindent {\em Proof of Eq.\  \Ref{nores}:} 

Eq.\ \Ref{Enp} implies
$$
E_{\vn}'-E_{\vn-\vmu}'  = \sum_j \mu_j [- \mu_j + 2n_j +\lambda(N+1-2j)]  
$$ 
with 
\eq
\label{muj}
\mu_j = (\vmu)_j = \sum_{k>j}\mu_{jk}- \sum_{k<j}\mu_{kj}.
\eqend 
Using 
\eq
\label{aj}
\sum_j \mu_j a_j = \sum_{j<k}\mu_{jk} (a_j-a_k)
\eqend
we get 
$$
E_{\vn}'-E_{\vn-\vmu}'
 = \sum_{j<k}\mu_{jk} [(\mu_k-\mu_j) + 2(n_j-n_k) +2\lambda(k-j)] 
$$
which proves Eq.\ \Ref{nores}. \QED

\section{The novel algorithm} 
This algorithm is based on the following Proposition which, roughly
speaking, is obtained by taking the Fourier transform of the
remarkable identity in Eq.\ \Ref{rem1} with respect to $\vy$.

\begin{Prop}\label{P1} Let $H$ be 
as in Eqs.\ \Ref{CS}--\Ref{V}. Then
\eq
\label{rem2}
H \hat F(\vx;\vn) = E_{\vn}\hat F(\vx;\vn)
- \ga \sum_{1\leq j<k\leq N}\sum_{\nu=1}^\infty 
\nu \hat F(\vx;\vn + \nu \vE_{jk}) 
\eqend
where 
\eq
\label{hF}
\hat F(\vx;\vn) = \cP_{\vn}(\vx) \Psi_0(\vx) \, ,\quad \vn\in\Z^N \eqend
with $\Psi_0$ as in Eqs.\ \Ref{Psi}--\Ref{psi} and 
\eq
\label{cP}
\cP_{\vn}(\vx) = \lim_{\eps\downarrow 0}
\int_{-\pi}^\pi \f{\dd y_1 }{(2\pi)}
\, \ee{\ii n_1  y_1 }\cdots  
\int_{-\pi}^\pi \f{\dd y_\ell }{(2\pi)}
\, \ee{\ii n_N  y_N }\, 
\f{ \prod_{1\leq j<k\leq N} \Bigl(1-\ee{\ii(
y_{j}-y_k )-(k-j)\eps} \Bigr)^{\lambda} } {\prod_{j,k=1}^N
\Bigl(1- \ee{\ii(x_k - y_j)-j\eps} \Bigr)^{\lambda} } 
\: ,  
\eqend
$\vE_{jk}$ as in Eq.\ \Ref{vE}, $E_{\vn}$ in Eq.\ \Ref{En}, 
and 
\eq
\label{ga}
\ga = 2\lambda(\lambda-1)\, . 
\eqend
\end{Prop}

\noindent {\it (Proof in Appendix~B.1.)} 

\smallskip

\begin{Rem}
To see that these functions $\cP_{\vn}$ are well-defined, we note that
they can be written as
$$
\cP_{\vn}(\vx) = \oint_{\cC_1}\f{\dd\xi_1}{2\pi\ii
\xi_1}\, \xi_1^{n_1} \cdots \oint_{\cC_N}\f{\dd\xi_N}{2\pi\ii \xi_N}\, 
\xi_N^{n_N} \f{\prod_{j<k}(1-\xi_j/\xi_k)^\lambda }{\prod_{j,k}(1-\ee{\ii
x_k}/\xi_j)^\lambda } 
$$
with integration paths $\cC_j: \xi_j = \ee{\eps j}\ee{\ii y_j}$,
$-\pi\leq y_j\leq \pi$, where $\eps>0$ is arbitrary.
\end{Rem}

We now show that this proposition provides a solution algorithm: Eq.\
\Ref{rem2} implies that the action of $H$ on the functions $\hat
F(\vx;\vn)$ is triangular, i.e.,
$H\hat F(\vx;\vn)$ is a linear combination
of functions $F(\vx;\vn+\vmu)$ with $\vmu\geq \vzero$. We thus can make 
the following ansatz for eigenfunctions,
\eq
\label{fn1}
f_{\vn}(\vx) = \sum_{\vmu\geq \vzero} a_{\vmu} F(\vx;\vn+\vmu) \, ,
\eqend
and then $H f_{\vn} =Ef_{\vn}$ implies 
$$
 \sum_{\vmu\geq \vzero} F(\vx;\vn+\vmu)
\left( 
\Bigl[E_{\vn+\vmu} - E \Bigr] a_{\vmu} - 
\ga \sum_{1\leq j<k\leq N} \sum_{\nu=1}^{\mu_{jk}}
 \nu a_{\vmu - \nu \vE_{jk} }
\right) = 0 . 
$$
We thus see that we get a solution of $Hf_{\vn}=E f_{\vn}$ is we set 
$E=E_{\vn}$ and determine the coefficients $a_{\vmu}$ by the following 
recursion relations,
\eq
\Bigl[E_{\vn+\vmu} - E_{\vn}\Bigr] a_{\vmu} =
\ga \sum_{1\leq j<k\leq N} \sum_{\nu=1}^{\mu_{jk}}
 \nu a_{\vmu - \nu \vE_{jk} } 
\eqend
which has triangular structure: we can set $a_{\vzero}=1$ 
(this fixes the normalization), and then the other $a_{\vmu}$
can be determined recursively in terms
of the $a_{\vmu'}$ where $\vmu'<\vmu$, at least if there is no 
resonance, i.e., if $E_{\vn+\vmu} - E_{\vn}$ does not vanish. 
This is true due to the following 

\begin{Lem} \label{Lem1}
\eq
\label{DelE}
E_{\vn+\vmu} - E_{\vn} = 
\sum_{j=1}^N \mu_j^2 + \sum_{1\leq j<k\leq N} 2\mu_{jk} \bigl[(n_j-n_k) + 
\lambda(k-j)\bigr]
\eqend
with $\mu_j$ in Eq.\ \Ref{muj}, which is manifestly positive provided that
Eq.\ \Ref{n1n2} holds true.
\end{Lem}

\noindent {\it (Proof in Appendix~B.2.)} 

\smallskip 

Moreover, the following Lemma 
shows that the $f_{\vn}$ are in fact symmetric polynomials,
i.e., a finite linear combination of the functions $S_{\vn}$ in 
Eq.\ \Ref{Sn}. 

\begin{Lem} \label{Lem2} The functions $\cP_{\vn}$ in Eq.\ \Ref{cP}
all are symmetric polynomials in the variables $z_j=\exp{(\ii x_j)}$
which are  non-zero only if
\eq
\label{bed}
n_j+n_{j+1} + \ldots + n_N \geq 0 \quad \forall j=1,2,\ldots N \, . 
\eqend
They can be written as
\eq
\label{p1}
\cP_{\vn}(\vx) = \sum_{\vm} p_{\vn,\vm} S_{\vm}(\vx)
\eqend
with $S_{\vm}(\vx)$ as in 
Eq.\ \Ref{Sn}, and the coefficients are  
\eq
\label{p2}
p_{\vn,\vm} = \mbox{$\sum ''$} 
\prod_{1\leq j'<k'\leq N}\prod_{j,k=1}^N 
\binom{\lambda}{\mu_{j'k'}}\binom{-\lambda}{ \nu_{j k}}
(-1)^{\mu_{j'k'}+ \nu_{j k} } \, 
\eqend
where the sum $\sum''$ is over all non-negative integers 
$\mu_{jk},\nu_{jk}$ restricted by the following $2N$ equations, 
\eq
\label{p3}
n_j = \sum_{\ell =1}^N \nu_{\ell j} + \sum_{\ell=1}^{j-1} \mu_{\ell j} - 
\sum_{\ell=j+1}^{N} \mu_{j\ell} , \quad 
m_j =  \sum_{\ell =1}^N \nu_{j \ell} 
\eqend
and $m_1\geq m_2\geq \ldots \geq m_N\geq 0$, 
implying in particular that there are only terms such that 
\eq
\label{p4}
\sum_{j=1}^N m_j = \sum_{j=1}^N n_j  . 
\eqend
\end{Lem}

\noindent {\it (Proof in Appendix~B.3.)} 

\smallskip 

Indeed, this Lemma implies the sum in Eq.\ \Ref{fn1} has only a finite
number of non-zero terms (since there is only a finite number of
$\vmu$ such that $\vn'=\vn+\vmu$ obeys all the conditions in Eq.\
\Ref{bed}), and thus the $f_{\vn}$ are a finite number of terms each
of which is a finite linear combination of functions $S_{\vn}$ in Eq.\
\Ref{Sn}.

\section{Conclusions} 

We can summarize our discussion in the previous Sections as follows. 

\begin{Prop} \label{P0}
For each $\vn\in\Z^N$ such that $n_1\geq n_2\geq \ldots \geq n_N\geq
0$, the standard algorithm reviewed in Sections~3 and the novel one
presented in Section~4 both yield an eigenfunction $f_{\vn}$ of the
Sutherland Hamiltonian $H$ in Eq.\ \Ref{CS}. In both cases, this
eigenfunction is of the form
$$
f_{\vn}(\vx) = \Phi_{\vn}(\vx)\Psi_0(\vx)
$$
with $\Psi_0$ in Eqs.\ \Ref{Psi}--\Ref{psi} and $\Phi_{\vn}$ a
symmetric polynomial in the variables $z_j=\exp{(\ii x_j)}$, and the
corresponding eigenvalues $E_{\vn}$ are given in Eq.\ \Ref{En}.
\end{Prop}

It thus follows from Theorem~3.1 in Ref.\ \cite{St} that, for
non-degenerate eigenvalues $E_{\vn}$, the eigenfunctions $f_{\vn}$
obtained with the two algorithms are equal (up to normalization),
and the functions $\Phi_{\vn}$ are proportional to the so-called Jack
polynomials (see Section~2 in Ref.\ \Ref{F} for details\footnote{The
latter Reference actually seems to suggest that this is true even for
non-degenerate eigenvalues.}).  We feel that this is quite remarkable
since, even though the two algorithms look somewhat similar and both
yield the same solution, there are several differences in details:

\begin{itemize}
\item The building blocks of the eigenfunctions in the
novel algorithm are the functions $\cP_{\vn}$ defined in Eq.\ \Ref{cP} and
not the plane waves $S_{\vn}$ in Eq.\ \Ref{Sn}.  
\item With the standard algorithm, one obviously obtains
eigenfunctions with polynomials $\Phi_{\vn}$ which have the form
$$
\Phi_{\vn} = \sum_{\vm\leq \vn} v_{\vn,\vm} S_{\vm}
$$
where the partial order here is defined as 
\eq \vm\leq \vn \; :
\Leftrightarrow \sum_{j=1}^k m_j \leq \sum_{j=1}^k n_j \quad \forall
k=1,2,\ldots N 
\eqend 
(this is called {\em dominance ordering} in \cite{St}; the latter fact
follows from Eq.\ \Ref{Phin} and $\vn\geq \vn-\vmu$ for all $\vmu\geq
\vzero$).  This is not at all obvious for the eigenfunctions obtained
with the novel algorithm (but of course should be true as well, at
least for non-degenerate eigenfunctions).
\item
In both algorithms it is important to rule out the occurrence of
resonances, but the reason for that is different (cf.\ Eq.\
\Ref{nores} with Lemma~\ref{Lem1} above, and observe the different
sign of $\vmu$).
\item
In the novel algorithm the restriction in Eq.\ \Ref{n1n2} can be
dropped, and in fact the solutions thus obtained are relevant
in the elliptic case \cite{EL0}.  There seems no way to drop this
restriction in the standard algorithm.
\item
From
Sutherland's algorithm it seems somewhat surprising that the
eigenvalues all can be written in the simple form $E_{\vn} = \sum_j
P_j^2$, but from the novel algorithm this is obvious. 
\item
As discussed in the Introduction, the novel algorithm can be
generalized to the elliptic case \cite{EL0,EL1}.
\end{itemize}

\bigskip

\noindent {\bf Acknowledgements:} I thank P.G.L.\ Mana for
suggestions on the manuscript. This work was supported by the Swedish
Natural Science Research Council (NFR).

\section*{Appendix A: Proof of Fact~2}
\app
We note that, 
\eq
F(\vx;\vy) = \Psi_0(\vx) \Psi_0(\vy) \f{1}{\prod_{j,k=1}^N 
\psi(x_j-y_k)^\lambda },
\eqend
with $\Psi_0$ in Eq.\ \Ref{Psi} and $\psi$ in Eq.\ \Ref{psi}. 
Using $H\Psi_0=E_0\Psi_0$ \cite{Su1} and the Leibniz rule of 
differentiation 
we obtain
\nonueqa
H(\vx)F(\vx;\vy) = \biggl(E_0 + 2\lambda^2 \sum_{j,\ell}\sum_{k \neq j}
\phi(x_j-x_k)\phi(x_j-y_\ell)+  \nonu 
+ \sum_{j,k}\lambda \phi'(x_j-y_k) 
- \sum_{j,k,\ell} \lambda^2  \phi(x_j-y_k)  \phi(x_j-y_\ell)  
\biggr)F(\vx;\vy)   
\nonueqaend
where 
\eq
\phi(r) = \psi'(r)/\psi(r) = \half\cot(r/2)
\eqend
(the prime indicates differentiation). Thus 
$$
\Bigl[ H(\vx)-H(\vy) \Bigr] F(\vx;\vy) =  \lambda^2 (\cdot)  F(\vx;\vy) 
$$
with 
$$
(\cdot) \equiv  
\biggl[ \sum_{j,\ell}\sum_{k \neq j} 2\phi(x_j-x_k)\phi(x_j-y_\ell) 
-  \sum_{j,k}\sum_{\ell\neq k} \phi(x_j-y_k)  \phi(x_j-y_\ell) \biggr] -
\biggl[\vx\leftrightarrow\vy\biggr]   
$$
where `$[\vx\leftrightarrow \vy]$' means `the same terms but with
$\vx$ and $\vy$ interchanged' (we used that all terms which are
even under $[\vx\leftrightarrow \vy]$ cancel). Relabeling
indices and using $\phi(r)=-\phi(r)$ we rewrite
\nonueqa
(\cdot) =  \sum_{j,\ell}\sum_{k \neq j}\biggl[\phi(x_j-x_k)\phi(x_j-y_\ell) 
+ \phi(x_k-x_j)\phi(x_k-y_\ell) - \nonu -  \phi(x_\ell-y_j)  \phi(x_\ell-y_k) 
\biggr] - \biggl[\vx\leftrightarrow\vy\biggr]  =  
  \sum_{j,\ell}\sum_{k \neq j}\biggl[\phi(x_j-x_k)\phi(x_j-y_\ell)
+  \nonu + \phi(x_j-x_k)\phi(y_\ell-x_k) +  \phi(y_\ell-x_j)  \phi(y_\ell-x_k)
 \biggr] - \biggl[\vx\leftrightarrow\vy\biggr] . 
\nonueqaend
We now can use the trigonometric identity 
\eq
\cot(x)\cot(y) +\cot(x)\cot(z) + \cot(y)\cot(z) = 1 
\quad \mbox{ if $x+y+z=0$} , 
\eqend
which shows that
$$
\phi(x_j-x_k)\phi(x_j-y_\ell)
+  \phi(x_j-x_k)\phi(y_\ell-x_k) +  \phi(y_\ell-x_j)  \phi(y_\ell-x_k) 
= -\half
$$
and thus proves $(\cdot)=0$. \QED

\section*{Appendix B: Other proofs}
\app
\subsection*{B.1 Proof of Proposition \ref{P1}}

We first observe two simple but useful facts. Firstly, 
the relation in Eq.\ \Ref{rem1} remains true if we replace $F(\vx;\vy)$
by
\eq
\label{Fp}
F'(\vx;\vy) = c \, \ee{ \ii P \sum_{j=1}^N (x_j-y_j)} \, F(\vx;\vy)
\eqend
for arbitrary constants $P\in\R$ and $c\in\C$. [To see this, introduce
center-of-mass coordinates $X=\sum_{j=1}^N x_j/N$ and $x'_j=(x_j-x_1)$
for $j=2,\ldots,N$, and similarly for $y$. Then
$H(\vx)= - \partial^2/\partial X^2 + H_c(\vx')$, and similarly for
$H(\vy)$. Invariance of Eq.\ \Ref{rem1} under $F\to \ee{-\ii P(X-Y)N}F$
thus follows from $(\partial/\partial X + \partial/\partial
Y)F(\vx;\vy)=0$, and the latter is implied by the obvious
invariance of $F(\vx;\vy)$ under $x_j\to x_j+a$, $y_j\to y_j+a$, 
$a\in\R$.  The invariance of Eq.\ \Ref{rem1} under $F\to c F$ is trivial,
of course].  Secondly, the variables $y_j$ in Eq.\
\Ref{rem1} need not be real but can be complex.

As mentioned, we intend to perform a Fourier transformation of the
identity in Eq.\ \Ref{rem1}, i.e.\, apply to it $(2\pi)^{-N}\int
\dd^N\vy\,\ee{\ii \vP\cdot \vy }$ with suitable momenta $\vP$. We
need to do this with care: firstly, the differential operator $H(\vy)$
has singularities at points $y_j=y_k$, and secondly, the function
$F(\vx;\vy)$ is not periodic in the variables $y_j$ but changes by
phase factors under $y_j\to y_j+2\pi$.  We therefore need to specify
suitable integration contours for the $y_j$'s avoiding the singular
points, and we need to choose the $P_j$ so as to compensate the
non-periodicity. To do that, we replace the real coordinates $y_j$ by
\eq 
z_j = y_j - \ii j \eps 
\eqend 
with $\eps>0$ a regularization parameter: as we will see, we can then 
integrate along the straight lines from $y_j=-\pi$ to $\pi$ and after
that perform the limit $\eps\downarrow 0$.  Since for all $j<k$,
$z_j-z_k =y_j-y_k+\ii\eps_{kj} $ with $\eps_{kj} = (k-j)\eps>0$, we
can use
\eq 
\label{sin}
\sin[(y+\ii\eps)/2] =\half
\ee{\ii\pi/2 } \ee{- \ii y/2 + \eps/2} 
\bigl(1- \ee{\ii y-\eps}\bigr) 
\eqend
for $\eps>0$. Taking the $\log$ of this identity and differentiating
we obtain $(1/2 \cot[(y+\ii\eps)/2]= -\ii /[1- \exp{(\ii y-\eps)}  ]$.
Expanding the r.h.s.\ in a geometric series and differentiating 
once more yields
\eq 
\label{sinsq}
\f{1}{4\sin^2[(y + \ii\eps)/2]} = 
-\sum_{\nu=1}^\infty \nu \ee{\ii\nu y-\nu\eps} . 
\eqend
This accounts for all
singularities and branch cuts in a consistent way.  
To determine the suitable $\vP$ use Eq.\ \Ref{sin} 
and compute 
\nonueqa
F(\vx;\vz) = (\cdots) \, \Psi_0(\vx)  \check \cP^\eps(\vx;\vy)
\nonueqaend 
with $\Psi_0(\vx)$ in Eqs.\ \Ref{Psi}--\Ref{psi}, 
\eq
\label{wcP}
\check \cP^\eps(\vx;\vy) = \f{ \prod_{1\leq j<k\leq N} \Bigl(1-\ee{\ii(
y_j-y_k )-(k-j)\eps} \Bigr)^{\lambda} } {\prod_{j,k=1}^N
\Bigl(1-\ee{\ii( x_j- y_{k})-k \eps} \Bigr)^{\lambda} } 
\eqend 
a function periodic in all the $y_j$, and 
\nonueqa 
(\cdots) =
\Bigl(\half\ee{\ii\pi\lambda/2}\Bigr)^{N(N-1)/2-N^2} \f{
\prod_{1\leq j<k\leq N} \ee{-\ii\lambda( y_{j}-y_k
)/2 + \lambda(k-j)\eps/2} } {\prod_{j,k=1}^N \ee{-\ii\lambda(x_j-y_k  
)/2 + \lambda k\eps/2}} \nonu
= const. \,  \ee{ \ii\lambda N \sum_{j=1}^N
(x_j-y_j)/2} \ee{ -\ii\lambda \sum_{j=1}^N (N+1-2j)y_j/2} 
\nonueqaend
(we used 
$\sum_{j<k}(y_j-y_k) = \sum_j (N+1-2j)y_j$). We thus see that
we can choose $P$ and $c$ in Eq.\ \Ref{Fp} such that
\eq
F'(\vx; \vz) = \ee{-\ii\lambda\sum_{j=1}^N [(N+1)/2-j]y_j} 
 \check \cP^\eps(\vx;\vy) \Psi_0(\vx) \: .   
\eqend
We need to choose the Fourier variables $\vP=(P_1,\ldots,P_N)$
such that $\ee{\ii \vP\cdot \vy}F'(\vx;\vz)$ is periodic in 
all $y_j$, and this implies 
\eq
\label{Pj}
P_{j} = 
n_j + \lambda[\half(N+1)-j], \quad n_j\in\Z .  
\eqend

We now can apply $(2\pi)^{-N}\int \dd^N\vy\,\ee{\ii \vP\cdot \vy} $
to the identity $H(\vx)F'(\vx;\vz) = H(\vz) F'(\vx;\vz)$. We recall 
$$
 H(\vz) = -\sum_j \f{\partial^2}{\partial y_j^2} + \ga
\sum_{j<k} \f{1}{4\sin^2[(y_j-y_k+\ii (k-j)\eps )/2]} 
$$
and use Eq.\ \Ref{sinsq}. After taking the limit $\eps\downarrow 0$ we
obtain Eq.\ \Ref{rem2}: the l.h.s.\ is obvious (note that $\hat F$ is
the Fourier transform of $F'$). The r.h.s.\ has two terms. The first
one is equal to $\sum_j P_j^2 \hat F$ and comes from the derivative
terms which we evaluated by partial integration. The second term is
obtained from the $1/\sin^2$-terms in $H(\vz)F'$ which we computed using
Eq.\ \Ref{sinsq}.\QED

\subsection*{B.2 Proof of Lemma \ref{Lem1}}

We write $(\vn+\vmu)_j = n_j+\mu_j$ with $\mu_j$ in Eq.\ \Ref{vmu}. 
Thus Eqs.\ \Ref{En} and \Ref{aj} imply, 
$$
E_{\vn+\vmu}-E_{\vn} = \sum_j \biggl( \mu_j^2 + 
2\mu_j (n_j + \lambda [\half(N+1)-j ] ) \biggr) ,
$$
and with Eq.\ \Ref{aj} we obtain Eq.\ \Ref{DelE}. \QED

\subsection*{B.3 Proof of Lemma \ref{Lem2}}
It is straightforward to evaluate $\cP_{\vn}(\vx)$
in Eq.\ \Ref{cP} by expanding all terms in Taylor series (using the
binomial series) and then performing the $y_j$ integrations which
corresponds to a projection onto the $y_j$-independent terms. 
The results is 
\eq
\cP_{\vn}(\vx) = \sum\mbox{$'$} 
\prod_{1\leq j'<k'\leq N}\prod_{j,k=1}^N 
\binom{\lambda}{\mu_{j'k'}}\binom{-\lambda}{ \nu_{j k}}
(-1)^{\mu_{j'k'}+ \nu_{j k} } \,  \ee{ \ii \nu_{j k }x_j } 
\eqend
where the sum $\sum'$ is over all non-negative 
integers $\mu_{kk'}$ and $\nu_{j\ell}$ such that
\eq
\label{coeff}
n_j - \sum_{\ell =1}^N \nu_{\ell j} - \sum_{\ell=1}^{j-1} \mu_{\ell j}
+ \sum_{\ell=j+1}^{N} \mu_{j\ell} =0 . \eqend 
Recalling the definition of $S_{\vn}$ in Eq.\ \Ref{Sn} we obtain
Eqs.\ \Ref{p1}--\Ref{p3}.

We now argue that this latter system of equations can have solutions only if
the conditions in Eq.\ \Ref{bed} all hold, which implies that
otherwise $\cP_{\vn}$ is zero. To see this we add up the last
$N+1-k$ relation in Eq.\ \Ref{coeff} ($k=N,N-1,\ldots,1$), and
by a relabeling of indices we obtain 
$$
\sum_{j=k}^N n_{j} = 
  \sum_{j=k}^N  
\sum_{\ell=1}^N \nu_{\ell j} + \sum_{\ell=1}^{k-1} \sum_{j=k}^N  \mu_{\ell j} 
$$
where the r.h.s.\ is always manifestly positive. This proves Eq.\
\Ref{bed}.  Setting $k=1$ and comparing with Eq.\ \Ref{p3} we
obtain Eq.\ \Ref{p4}. Moreover, for fixed $n_j$, there are at most a
finite number of different solutions of Eq.\ \Ref{coeff}, implying
that $\cP_{\vn}$ is a polynomial.  To see that we write Eq.\ \Ref{coeff} as
follows, \eq
\label{coeff1}
n_j + \sum_{\ell=j+1}^{N} \mu_{j\ell} 
 = \sum_{\ell =1}^N \nu_{\ell j} + \sum_{\ell=1}^{j-1} \mu_{\ell j}  
\eqend
and determine possible solutions for decreasing values of $j$ starting
at $j=N$. It is easy to prove by induction that there is only a finite
number of solutions 
$$\bigl\{ \nu_{\ell j}\bigr\}_{j, \ell=1}^N,
\bigl\{ \mu_{j\ell}\bigr\}_{1\leq j<\ell\leq N} \in\N_0^{N^2+N(N-1)/2}$$
of this system of equations: For $j=N$ we get 
$$
n_N
 = \sum_{\ell =1}^N \nu_{\ell N} + \sum_{\ell=1}^{N-1} \mu_{\ell N}  
$$ 
and there is obviously only a finite number of different solutions
$\bigl\{ \nu_{\ell N}\bigr\}_{\ell=1}^N ,\bigl\{ 
\mu_{\ell N}\bigr\}_{\ell=1}^{N-1} $ of that equation. If we
consider Eq.\ \Ref{coeff1} for some $j=j_0<N$, the possible solutions  
for $\bigl \{ \mu_{j_0,\ell} \bigr\}_{\ell>j_0}$ were
already determined by the equations for $j>j_0$ and, by the induction
hypothesis, there is only a finite number of them. One thus only has to
consider a finite number of equations, and each of them obviously has
only a finite number of solutions 
$\bigl\{ \nu_{\ell j}\bigr\}_{\ell=1}^N ,\bigl\{ 
\mu_{\ell j}\bigr\}_{\ell=1}^{j-1} $.\QED

\end{document}